\documentclass[aps,showpacs,amsmath,amssymb,12pt]{revtex4}
\usepackage{latexsym}
\usepackage{bm}

\def\G{\Gamma}
\def\a{\alpha} 
\def\b{\beta}
\def\d{\delta} 
\def\l{\lambda} 
\def\m{\mu} 
\def\n{\nu}
\def\s{\sigma}
\def\p{\phi}

\def\vp{\varphi}
\def\r{\rho} 
\def\th{\theta}

\def\be{\begin{equation}}
\def\ee{\end{equation}}

\def\dd{\partial}

\def\ep{\epsilon}

\begin{document}

\title{Topologically massive gravity as a Pais-Uhlenbeck oscillator} 

\author{{\"O}zg{\"u}r Sar{\i}o\u{g}lu}
\email{sarioglu@metu.edu.tr}
\affiliation{Department of Physics, Faculty of Arts and Sciences,\\
             Middle East Technical University, 06531, Ankara, Turkey}

\author{Bayram Tekin}
\email{btekin@metu.edu.tr}
\affiliation{Department of Physics, Faculty of Arts and Sciences,\\
             Middle East Technical University, 06531, Ankara, Turkey}

\date{\today}

\begin{abstract} 
We give a detailed account of the free field spectrum and the Newtonian 
limit of the linearized ``massive'' (Pauli-Fierz), ``topologically massive'' 
(Einstein-Hilbert-Chern-Simons) gravity in $2+1$ dimensions about a 
Minkowski spacetime. For a certain ratio of the parameters, the linearized 
free theory is Jordan-diagonalizable and reduces to a degenerate 
``Pais-Uhlenbeck" oscillator which, despite being a higher derivative 
theory, is ghost-free.   
\end{abstract}


\maketitle


\section{\label{intro} Introduction}

As is well known, Einstein's General Relativity (GR) in 3+1 dimensions
is nonrenormalizable and breaks down at ``high'' energies. If GR is
considered as a low energy effective limit of some unknown fundamental
theory, then it will receive higher curvature corrections, such as
\( \a R^2 + \b R_{\m\n}^2 \), etc. which give a much well-behaved gravity
theory in the UV region, but unfortunately ghosts
are not decoupled and unitarity is lost. In fact, in the Einstein plus 
quadratic curvature model whose Newtonian potential reads
\[ V(r) = G M \Big\{ - \frac{1}{r} - \frac{1}{3} \frac{e^{- m_1 r}}{r}
+ \frac{4}{3} \frac{e^{- m_2 r}}{r} \Big\} \, , \]
one can see the nice UV behavior ($V(0)=0$) and the presence of the
ghost related to the last repulsive term (see \cite{von} and the early
references therein). It is, therefore, of extreme importance to get a
higher derivative gravity theory without ghost problems. 2+1 dimensional
gravity, which is quite easy compared to GR, is a nice laboratory to
study various quantum gravity issues. Admittedly, the lesson one learns
for 3+1 dimensions should always be taken with a grain of salt. Nevertheless,
valuable insight on higher derivative gravity theories, as well as massive
gravity theories, can be gained with the study of 2+1 dimensional gravity
models. In fact, an analogous approach in quantum field theory has borne
much fruit. In this paper, we shall find a particular ghost-free higher
derivative gravity model, which has various ``mass'' terms.

In $2+1$ dimensions, in addition to the usual
gauge non-invariant Proca/Pauli-Fierz mass terms, one can add a gauge 
invariant ``mass" which is the Chern-Simons term \cite{djt} 
to both (the abelian and the non-abelian) spin-1 and spin-2 models.
[To be more precise, the Chern-Simons term acts like a mass term only 
at the quadratic level.] There are major differences between the vector 
and the tensor cases, one of which is related to the gauge invariance: 
In the former, even the Proca mass can be made gauge invariant by introducing 
a Higgs field [see e.g. \cite{pth} and the references therein]; but in the 
latter, it is simply not known how to get the Pauli-Fierz mass from a gauge 
(diffeomorphism) invariant action.   

The van Dam-Veltman-Zakharov (\cite{dv,zak}) discontinuity between the 
massive and massless gravity models registers as another difference between 
the {\em vector} and {\em tensor} theories. [Recall that the vDVZ 
discontinuity states that around the Minkowski background, the massless limit 
of the massive gravity (Einstein-Pauli-Fierz theory) is not the 
massless gravity (Einstein theory).] The existence of such a 
discontinuity \footnote{According to Vainshtein \cite{vain}, the 
discontinuity should disappear once non-perturbative corrections are taken 
into account: Namely, in addition to the mass of the graviton, dimensionful
scales, related to the Schwarzschild radius of, say, one of the 
scattering particles should become relevant. This is a plausible idea since 
only then, the small graviton mass limit seems to make sense, however we are 
not aware of a full demonstration of Vainshtein's conjecture.} and perhaps the 
isolation of massless gravity from the massive (albeit arbitrarily small) 
ones, is of extreme importance as far as the concept of mass in gravity is 
concerned. The discontinuity also has much to do with the difference in the 
degrees of freedom between the massive and massless models.   

In a previous Letter \cite{dt}, the 3-term model (Einstein-Hilbert, 
Chern-Simons and Pauli-Fierz) was studied in the context of various 
discontinuities and parameter ranges. The degrees of freedom were 
identified but the diagonalization of the model (or the explicit 
form of the free field Lagrangian) was not carried out. As we shall see 
in this current work, a detailed analysis of the free theory is not 
redundant at all, since the theory exhibits rather interesting 
features: For example, as opposed to being a collection of simple harmonic 
oscillators, we shall get higher order {\em degenerate} (i.e. equal 
frequency, mass) Pais-Uhlenbeck oscillators \cite{pu} and the 
model (for a certain ratio of the Chern-Simons and the Pauli-Fierz 
parameters) can be brought into a Jordan-diagonal form at most. 
This is rather unexpected since, without the Pauli-Fierz mass term, the 
free field limit of the model is {\em not} a higher derivative model. 
Thus we have a very interesting situation here: the higher derivative model 
is ghost-free! In fact, in the recent literature, one can find examples of 
similar ghost-free higher derivative models: Mannheim-Davidson \cite{md}, 
Smilga \cite{sm1, sm2} and Hawking-Hertog \cite{hh} all deal with the 
one-dimensional degenerate Pais-Uhlenbeck oscillator as a model of higher 
derivative ghost-free theories. 

There are a number of reasons why one would be interested in higher 
derivative models: First of all, they are much better behaved in the UV 
region (see e.g. Stelle's \cite{st} work in higher curvature gravity) 
and secondly, even if one does not have higher order terms at the 
classical level, quantum corrections usually introduce such terms. 
Ironically, in most of the renormalizable theories, during the 
process of regularization and renormalization, one introduces ghosts only 
to remove them at the end! Thus a higher derivative model without ghosts, 
as the one considered here, would be of some interest.

The layout of the paper is as follows: In section 2, we introduce the 
3-term model and work out the free field Lagrangian. In section 3, we 
suppress the spatial dependencies to basically look at the model as a 
collection of oscillators. In the first appendix, we give a basic review
of the Pais-Uhlenbeck oscillator and the second appendix contains the 
Newtonian limit of the 3-term model.

\section{\label{model} The model}

Since the signs and the values (or the ratios) of the involved parameters 
are of extreme importance, here we carefully define and work out the linear 
version of the 3-term gravitational action which is the sum of 
Einstein-Hilbert, (third derivative order) Chern-Simons and standard
Pauli-Fierz mass, terms
\be
I = \int_M d^3 x \, \left\{ a \sqrt{-g} R  - \frac{1}{2 \mu} 
\ep^{\l\m\n} \G^{\r}\,_{\l\s} \left( \dd_\m \G^{\s}\,_{\r\n} + \frac{2}{3}
\G^{\s}\,_{\m\b} \G^{\b}\,_{\n\r} \right) - \frac{m^2}{4} \sqrt{-g} 
( h_{\m\n} h^{\m\n} - h^2 ) \right\} \, ,
\label{fullaction}
\ee
only at quadratic order in deviations 
\( h_{\m\n} \equiv g_{\m\n} - \eta_{\m\n} \) about the flat Minkowski
background $\eta_{\m\n}$ with the usual definition 
\( h \equiv \eta^{\m\n} h_{\m\n} \). Our signature is $(-,+,+)$, 
we use $\ep_{012} \equiv 1$ throughout and, in what follows, all 
operations are carried out with respect to the flat background 
$\eta_{\m\n}$. The model under investigation is the most general such
model \cite{dt} since the sign of $\m$ is arbitrary, that of $m^2$ is
free and the presence of $a$ allows for choosing the Einstein term's
sign ($a=+1$ in the usual case, $a=-1$ for topologically massive gravity
(TMG) \cite{djt} and $a=0$ for CS-Pauli-Fierz theory). Note that 
$\m \to \infty$ represents massive gravity with 2 excitations and 
$m = 0$ is TMG with a single mode \cite{djt}, whereas pure Einstein 
theory, $1/ \m = m^2 =0$, has no excitations in $D=3$. Without working 
the details, a cursory look at the model does not really reveal what 
kinds of ``oscillators'' we have for the free field limit. In fact, as 
will be seen shortly, there are various discontinuities as far as the 
limits of the parameters are concerned. 
 
We shall be interested only with the free field limit for now. [Of course, 
one eventually has to deal with the interacting theory, but, 
unfortunately, there are infinitely many terms in the interaction 
Lagrangian rendering the perturbation theory intractable.] 
The individual terms that make up the action (\ref{fullaction}) can 
each be expanded at quadratic order in $h_{\m\n}$ as follows:
The Einstein-Hilbert piece reads
\be 
I_{E} = a \int d^3 x \, \sqrt{-g} R = - \frac{a}{2} \int d^3 x \,
h_{\m\n} \, {\cal{G}}_{L}^{\m\n} + O(h^3) \, ,
\label{einst}
\ee 
where 
\be
{\cal{G}}_{L}^{\m\n} = \frac{1}{2} \left( \dd_{\s} \dd^{\m} h^{\n\s}
+ \dd_{\s} \dd^{\n} h^{\m\s} - \Box h^{\m\n} - \dd^{\m} \dd^{\n} h 
\right) - \frac{1}{2} \eta^{\m\n} \left( \dd_{\a} \dd_{\b} h^{\a\b}
- \Box h \right) \, .
\ee
The Chern-Simons bit yields
\be 
I_{CS} = - \frac{1}{2 \m} \int d^3 x \, \ep^{\l\m\n} \G^{\r}\,_{\l\s} 
\left( \dd_\m \G^{\s}\,_{\r\n} + \frac{2}{3} \G^{\s}\,_{\m\b} 
\G^{\b}\,_{\n\r} \right) = - \frac{1}{2 \m} \int d^3 x \,
\ep_{\m\a\b} \, {\cal{G}}_{L}^{\a\n} \, \dd^{\m} h^{\b}\,_{\n} + O(h^3) \, .
\label{chsim}
\ee
Finally the Pauli-Fierz mass term, up to $O(h^3)$, gives
\be 
I_{PF} = - \frac{m^2}{4} \int d^3 x  \, \sqrt{-g}  
( h_{\m\n} h^{\m\n} - h^2 ) = - \frac{m^2}{4} \int d^3 x  \, 
( h_{ij}^2 - 2 h_{0i} h_{0i} + h_{00}^2 - h^2 ) \, . 
\label{pafie}
\ee

It was shown in \cite{djt} that the sign of the Einstein term in pure 
TMG ($m^2 = 0$) must be $a= -1$, opposite to that in the usual Einstein 
gravity, in order for the energy to be positive, independent of the sign of
$\m$. On the other hand, the usual massive spin-2 system does have
excitations and this forces both the relative and overall signs of 
the Einstein and Pauli-Fierz mass terms to be in the usual way 
to avoid the presence of ghosts and tachyons. This indicates an
inevitable dilemma in the choice of the sign of $a$ in the TMG and 
massive spin-2 cases.

To analyze the full system given in (\ref{fullaction}) for generic 
values of ($\mu , m$) and its dependence on the sign of $a$, we first 
decompose $h_{\m\n}$ as
\be 
h_{ij} \equiv ( \d_{ij} + {\hat{\dd}}_i {\hat{\dd}}_j ) \p 
- {\hat{\dd}}_i {\hat{\dd}}_j \chi +
( \ep_{ik} \hat{\dd}_k \hat{\dd}_j +
\ep_{jk} \hat{\dd}_k \hat{\dd}_i ) \xi \, , \qquad
h_{0i} \equiv - \ep_{ij} {\dd}_j \eta + {\dd}_i N_{L} \, , \qquad
h_{00} \equiv N \, , 
\label{decom}
\ee 
where \( {\hat{\dd}}_i \equiv \dd_i / \sqrt{-{\dd}_{k}^{2}} \).  
The decomposition (\ref{decom}) yields the following for the
Einstein (\ref{einst}), Chern-Simons (\ref{chsim}) and Pauli-Fierz
(\ref{pafie}) components of (\ref{fullaction}), respectively: 
\begin{eqnarray} 
I_{E} & = & \frac{a}{2} \int d^3 x \, \left\{ \p \ddot{\chi} + 
\p \nabla^2 ( N - 2 \dot{N_{L}} ) + ( - \nabla^2 \eta + \dot{\xi} )^2 
\right\} \, , \\
I_{CS} & = & \frac{1}{2 \m} \int d^3 x \, \left\{ ( - \nabla^2 \eta
+ \dot{\xi} ) [ \nabla^2 ( N - 2 \dot{N_{L}} ) + \ddot{\chi} +
\Box{\p} ] \right\} \, , \\
I_{PF} & = & - \frac{m^2}{2} \int d^3 x \, \left\{ N_{L} \nabla^2 N_{L} +
\eta \nabla^2 \eta + \xi^2 - \p \chi + N (\p + \chi) \right\} \, .
\end{eqnarray}
When all these are put together, the terms proportional to $N$ can be
collected altogether to obtain
\begin{eqnarray} 
I & = & \frac{1}{2} \int d^3 x \, \left\{ a ( \p \ddot{\chi} 
- 2 \p \nabla^2 \dot{N_{L}} + ( - \nabla^2 \eta + \dot{\xi} )^2 ) + 
\frac{1}{\m} ( - \nabla^2 \eta + \dot{\xi} ) [-2 \nabla^2 \dot{N_{L}}  
+ \ddot{\chi} + \Box{\phi} ]  \right. \nonumber \\
& & \hspace{-0.5cm}
\left. -m^2 [N_{L} \nabla^2 N_{L} + \eta \nabla^2 \eta + \xi^2 - \p \chi] 
+ N \left[ a \nabla^2 \p - \frac{1}{\m} \nabla^2 ( \nabla^2 \eta - \dot{\xi} ) 
- m^2 (\p + \chi) \right] \right\} \, . \label{Imtmg}
\end{eqnarray}
This looks like a highly complicated system. To simplify it, the next 
thing to do is to use integrations by parts, whenever necessary,
to eliminate those variables from (\ref{Imtmg}) which can be treated as
`Lagrange multipliers' and to use the corresponding `constraint equations'
thus obtained so that only dynamical degrees of freedom are left. When $N$,
$\eta$ and $N_{L}$ are eliminated in this fashion, one finds 
\be
I = \frac{1}{2} \int d^3 x \, \left[ a \p \Box \p - m^2 \p^2 + 2 \l \Box \p
+ a \m^2 \l^2 - m^2 (\m \l - \dot{\xi}) \frac{1}{\nabla^2} (\m \l - \dot{\xi})
- m^2 \xi^2 \right] \, ,
\ee
where we have used \( \l \equiv - a \p + m^2 \frac{1}{\nabla^2} (\p + \chi) \)
for simplicity. The remaining three dynamical components of $h_{\m\n}$ yield
\begin{eqnarray}
\d \p: \quad & & \Box \l + a \Box \p - m^2 \p = 0 \, , \\
\d \l: \quad & & \Box \p - m^2 \m^2 \frac{1}{\nabla^2} \l + m^2 \m 
\frac{1}{\nabla^2} \dot{\xi} + a \m^2 \l = 0 \, , \\
\d \xi: \quad & & \m \dot{\l} + \Box \xi = 0 \, ,
\end{eqnarray}
which together give
\be
(\Box^3 - a^2 \m^2 \Box^2 + 2 a m^2 \m^2 \Box - m^4 \m^2) \p = 0 \, ,
\label{phieqn} 
\ee
for the field $\p$, through which the remaining fields are determined as
\[ \l = \left( -a + m^2 \frac{1}{\Box} \right) \p \qquad \mbox{and} \qquad
\xi = \frac{1}{\Box} \left( a \m - m^2 \m \frac{1}{\Box} \right) \dot{\p} 
\, . \]
Equation (\ref{phieqn}) was obtained in \cite{dt} and \cite{pinheiro}, 
in which it was shown that for $a=1$, the roots of the eigenvalue 
equation are complex unless $\m^2/m^2 \ge 27/4$.

Introducing auxiliary variables \( \psi \equiv \Box \p \) and
\( \Omega \equiv \Box^2 \p = \Box \psi \), the equation for $\p$ 
(\ref{phieqn}) can be put in the form \( \Box {\bf x} = {\bf A x} \), where 
\[ {\bf x} = \left[
\begin{array}{c}
  \p \\
  \psi \\
  \Omega 
\end{array}
\right] \, , \qquad
{\bf A} = \left[
\begin{array}{ccc}
  0 & 1 & 0 \\
  0 & 0 & 1 \\
  m^4 \m^2 & -2 a m^2 \m^2 & a^2 \m^2 
\end{array}
\right] \, . \]

In passing, we should note that when $\m^2/m^2 > 27/4$, the eigenvalue
equation has three distinct real roots, the explicit forms of which are 
rather cumbersome so that we refrain from presenting them here. However, 
the free field limit of the model can simply be thought of as being 
described by three uncoupled oscillators then. 

{\bf {The degenerate case:}}

Let us now consider the case when $a=1$ and $\m^2/m^2 = 27/4$. Then 
\( \det{(p {\bf I} - {\bf A})} = (p - 3 m^2)^2 \, (p - \frac{3}{4} m^2) \) 
and by using the eigenvectors of ${\bf A}$, one can form the modal matrix 
of ${\bf A}$ as
\[ {\bf P} = \left[
\begin{array}{ccc}
  16/9 & 1 & 0 \\
  4 m^2/3 & 3 m^2 & 1 \\
  m^4 & 9 m^4 & 6 m^2 
\end{array}
\right] \, , \]
which `Jordan-diagonalizes' ${\bf A}$ as
\[ {\bf \Lambda} = {\bf P}^{-1} {\bf A P} = \left[
\begin{array}{ccc}
  3 m^2/4 & 0 & 0 \\
  0 & 3 m^2 & 1 \\
  0 & 0 & 3 m^2 
\end{array}
\right] \, . \]
By introducing \( {\bf z} = {\bf P}^{-1} {\bf x} = [z_1 \; z_2 \; z_3]^{T} \),
this system thus takes the form \( \Box {\bf z} = {\bf \Lambda z} \),
which can be thought of as coming from an equivalent dynamical system
whose action reads
\[ I_{\rm{eq}} = \frac{1}{2} \int d^3 x \, \left[ z_1 \left( \Box - 
\frac{3}{4} m^2 \right) z_1 + z_2 (\Box - 3 m^2)^2 z_2 \right] \, , \]
where specifically
\[ z_1 = \left( 1 - \frac{2}{3 m^2} \Box + \frac{1}{9 m^4} \Box^2 \right) \p
\quad \mbox{and} \quad
z_2 = \left( - \frac{7}{9} + \frac{32}{27 m^2} \Box - \frac{16}{81 m^2} 
\Box^2 \right) \p \, . \]

The first term represents the usual (ghost-free) massive real scalar field 
whose quantization is well-known. The second term describes a degenerate 
(or purely quadratic) Pais-Uhlenbeck oscillator. In principle, just like 
any other higher derivative model, one expects this model to be contaminated 
with ghosts. However, as was shown in \cite{md, hh, sm1}, this is not true: 
Purely quadratic models differ from those with quadratic plus lower 
derivative ones. This should not be surprising because, even the classical 
solutions of the purely quadratic theories have milder instabilities and 
physically acceptable parameter ranges. 

This, of course, is all at the level of the free-field theory. When 
interactions are introduced, the picture will necessarily change drastically. 
We are not aware of much work on this, save the work of Smilga \cite{sm2} 
who has shown that once a quartic interaction is introduced, depending on 
the numerical values of the ratios of the involved parameters (such as 
the coupling constant to mass ratio), the model still has ghost-free, 
stable regions. This issue obviously deserves more attention. 

\section{\label{newt} The reduced models}

In this section, we shall suppress the spatial part of the theory to see 
more transparently the degrees of freedom (the free oscillators) of 
the model. The advantage of this is that (\ref{Imtmg}) simplifies a 
great deal without losing any of its degrees of freedom. Much of what we 
have done above (in a fully relativistic way) can be understood more 
easily from the following reduced Lagrangian 
\be
I = \frac{1}{2} \int d^3 x \, \left\{ a ( \p \ddot{\chi} + \dot{\xi}^2 ) 
+ \frac{1}{\m} \dot{\xi} \, ( \ddot{\chi} - \ddot{\p} ) 
- m^2 ( \xi^2 - \p \chi + N  (\p + \chi) ) \right\} \, .
\ee
Getting rid of the Lagrange multiplier $N$, we find the reduced form of the 
3-term model:
\be
I = \frac{1}{2} \int d^3 x \, \left\{ a ( \dot{\chi}^2  + \dot{\xi}^2 )
+ \frac{2}{\m} \, \dot{\xi} \, \ddot{\chi} - m^2 ( \xi^2 +\chi^2 ) 
\right\}
\, ,
\ee
which leads to the following equations of motion:
\be
\ddot{\chi} + m^2 \chi = \frac{1}{\m} \dddot{\xi} \,, \qquad \qquad
\ddot{\xi} + m^2 \xi = - \frac{1}{\m} \dddot{\chi} \, .
\ee
Let us now check various limits:

{\bf {The pure TMG case:}}

Setting the Pauli-Fierz mass to zero ($m^2 = 0$) and eliminating
$\dot{\xi}$, which behaves like a Lagrange multiplier, we have   
\be
I = \frac{1}{2} \int d^3 x \, \left\{ -a ( \dot{\chi}^2  
- a^2 \m^2 \chi^2 ) \right\} \, .
\label{3term}
\ee
It is clear that when $a = -1$, we have a ghost-free, massive, single 
excitation. This is the good old result of \cite{djt}.  

{\bf {Einstein-Pauli-Fierz case:}}

Setting $\m \to \infty$, we get 
\be 
I = \frac{1}{2} \int d^3 x \, \left\{ a ( \dot{\chi}^2  + \dot{\xi}^2 ) 
- m^2 ( \xi^2 +\chi^2 ) \right\} \, ,
\ee
which describes two real scalar, massive degrees of freedom if $a=+1$. 
[This is what one expects from the onset, since massive spin-2 field has 
2 degrees of freedom in 2+1 dimensions, that is as much as the massless 
spin-2 field in 3+1 dimensions.]

{\bf {No Einstein term:}}

Setting $a=0$ yields a higher derivative model
\be
I = \frac{1}{2} \int d^3 x \, \left\{ \frac{2}{\m} \, \dot{\xi} \, 
\ddot{\chi} - m^2 ( \xi^2 +\chi^2 ) \right\} \, ,
\ee
which has to be diagonalized. However, as the eigenvalue equation 
\( (M^3 + |\m| m^2= 0) \), where $M$ is the mass of the excitations, 
constructed from the equations of motion, shows, this model has 
tachyonic solutions. This could be of some interest in Anti-de Sitter 
spacetime backgrounds which allows for certain negative mass squared 
solutions, but is not viable in flat backgrounds. 

\section{\label{conc} Conclusions}

In this work, we have studied massive gravity (with two different types 
of mass: the Chern-Simons and the Pauli-Fierz) in 2+1 dimensional gravity. 
The theory has interesting mass discontinuities: It becomes a higher order 
(degenerate) Pais-Uhlenbeck oscillator if the ratio of the Chern-Simons and 
the Pauli-Fierz masses is tuned to  $3 \sqrt{3} /2$. The full theory (i.e. 
the Einstein-Chern-Simons-Pauli-Fierz model) generically has, at the 
linearized level, 3 distinct massive degrees of freedom. The free field 
Lagrangian is a collection of these 3 oscillators. Therefore, it was 
surprising to see that when the parameters are tuned as 
mentioned above, two of the masses coalesce and one gets a higher 
derivative oscillator in addition to a normal one. The model at that limit 
is at most Jordan-diagonalizable. It is a ghost-free higher-derivative 
model whose properties were put under scrutiny recently \cite{sm1, md, hh}, 
where quantization was also carried out.     

In this work, we have only considered the non-interacting theory. It would 
be quite interesting to consider the lowest order interacting theory to 
see whether the higher derivative model stays ghost-free (or as far as the 
classical solutions are concerned, whether there are stable regions or not). 
Such an analysis was carried out by Smilga \cite{sm2} in a simplified 
model.   

\appendix
\section{\label{appa} The Pais-Uhlenbeck oscillator}

Here we would like to recapitulate the nonrelativistic limit of the
higher derivative field theories considered a long time ago by \cite{pu}.
Pais and Uhlenbeck studied a generic scalar field theory model whose 
Lagrangian is of the form
\[ {\cal L} = - \frac{1}{2} \p \Big( \prod_{i=1}^{N} (\Box - M_{i}^2) 
\Big) \p \, . \]
Here we will take $N=2$ and examine only the nonrelativistic limit,
which basically means the dropping down of the spatial derivatives.
Thus consider a Lagrangian of the form
\be
L = \frac{1}{2} \Big\{ \ddot{q}^2 - (\omega_{1}^2 + \omega_{2}^2) \dot{q}^2
+ \omega_{1}^2 \omega_{2}^2 q^2 \Big\} \, , \label{lag}
\ee
and note that \( \omega_{1} \neq \omega_{2} \) and 
\( \omega_{1} = \omega_{2} \) cases differ drastically. We will call
the latter case as the degenerate Pais-Uhlenbeck oscillator which is
similar to the case studied in this paper.

The naive Ostrogradski Hamiltonian simply reads
\be
H = \frac{\ddot{q}^2}{2} - \dot{q} \, \dddot{q} 
- \frac{1}{2} (\omega_{1}^2 + \omega_{2}^2) \dot{q}^2
- \frac{1}{2} \omega_{1}^2 \omega_{2}^2 q^2 \, ,
\ee
which gives the following, not necessarily positive, energies for solutions
of the equation of motion
\be 
E = \left\{
\begin{array}{lr}
\frac{1}{2} (\omega_{1}^2 - \omega_{2}^2) \Big[ \omega_{1}^2 (c_{1}^2 
+ c_{2}^2) - \omega_{2}^2 (c_{3}^2 + c_{4}^2) \Big], 
& \omega_{1} \neq \omega_{2} \\
2 \omega^2 \Big[ \omega (c_{5} c_{6} - c_{7} c_{8}) + c_{5}^2 + c_{8}^2
 \Big], & \omega_{1} = \omega_{2} = \omega
\end{array}
\right. \, ,
\ee
where $c_{i} (i=1, \dots, 8)$ are constants. To be able to quantize these
theories, a proper Hamiltonian in terms of momenta can be defined by using
Dirac's constraint analysis. Taking $\dot{q} = \pi$, one gets
\be
H = \frac{p_{\pi}^2}{2} + \frac{1}{2} (\omega_{1}^2 + \omega_{2}^2) \pi^2
- \frac{1}{2} \omega_{1}^2 \omega_{2}^2 q^2 + p_{q} \pi \, ,
\ee
where $p_{\pi}$ and $p_{q}$ are the usual canonical momenta. One can see
the glimpses of two coupled oscillators here, but one still needs to
``diagonalize'' this Hamiltonian. For $\omega_{1} = \omega_{2}$, just
like the oscillator studied in the bulk of this paper, the Hamiltonian is
only Jordan diagonalizable. For $\omega_{1} \neq \omega_{2}$, one can
use the following Pais-Uhlenbeck variables
\be
Q_{1} \equiv q + \frac{\ddot{q}}{\omega_{2}^2} \, , \qquad
Q_{2} \equiv q + \frac{\ddot{q}}{\omega_{1}^2} \, .
\ee
In this case the Hamiltonian becomes
\be
H = \frac{1}{2} \frac{\omega_{1}^4}{\omega_{2}^2 - \omega_{1}^2}
(\dot{Q}_2^2 + \omega_{2}^2 Q_{2}^{2})
- \frac{1}{2} \frac{\omega_{2}^4}{\omega_{2}^2 - \omega_{1}^2}
(\dot{Q}_1^2 + \omega_{1}^2 Q_{1}^{2}) \, ,
\ee
which is a collection of two oscillators except that one has the wrong sign.
For a discussion of the quantization of the $\omega_{1} \neq \omega_{2}$ case,
we refer the reader to \cite{md, sm1}.

\section{\label{appb} The solutions}

Let us now examine the steady-state solutions of the $D=3$ Pais-Uhlenbeck
oscillator. For this purpose, we assume that the mass parameter $m$ has
been scaled appropriately and the field $\p$ (or $z_2$ thereof) has been
redefined accordingly such that the problem reduces to investigating the
solutions of the \( (\nabla^2 - m^2)^2 \vp = 0 \) equation on the
(2-spatial dimensional full) plane. Below we will give the exact solution
of this equation in terms of highly complicated special functions. However,
one can easily extract the asymptotic behavior of the solution at spatial
infinity, which is simply \( \vp \sim (c_1 + c_2 r) e^{- m r} \). Of course,
we have assumed ``nice'' boundary conditions and discarded the divergent 
parts to find this. 

First consider the simpler equation \( (\nabla^2 - m^2) \vp = 0 \). For
the problem at hand, one can assume circular symmetry, take 
\( \vp = \vp(r, \th) \), employ a `separation of variables' ansatz, 
introduce a new independent variable \( x \equiv m r \), arrive at 
the well known modified Bessel equation and, further assuming the 
solutions to be free of any angular dependency, write down
\[ \vp(r, \th) = R (m r) \equiv R(x) = \a \, I_{0}(x) + \b \, K_{0}(x) \, , \]
for some integration constants $\a$ and $\b$. Thus, our original problem
reduces to examining the solutions of the following ordinary differential
equation
\[ \frac{d^2 R}{dx^2} + \frac{1}{x} \frac{dR}{dx} 
- R(x) = \b \, K_{0}(x) \, , \]
if we further demand that the solutions of \( (\nabla^2 - m^2)^2 \vp = 0 \)
are finite and regular at large $x$ (or $r$). One can view this as a 
two-point boundary condition problem, where the solution $R(x)$ is
required to be finite and bounded for both small and large values of $x$.
Using standard Green's functions' techniques, one arrives at
\[ R(x) = - \b \left( \int_{0}^{x} K_{0}(x) \, I_{0}(y) \, K_{0}(y) \, y \, dy 
+ \int_{x}^{\infty} I_{0}(x) \, K_{0}^{2}(y) \, y \, dy \right) \, , \]
which can be evaluated using the software package {\tt Mathematica}. The
outcome of this is
\[  R(x) = \vp(m r) = - \b \left( \frac{1}{2} \, x^2 \, I_{0}(x) 
( - K_{0}^{2}(x) + K_{1}^{2}(x) ) + \frac{1}{4 \sqrt{\pi}} \, K_{0}(x) 
\, u(x) \right) \, , \]
where
\[ u(x) = {\rm MeijerG}[\{ \{ 3/2 \}, \{ \, \} \}, 
\{ \{ 1,1 \} , \{ 0 \} \} \}, x^2] =
\frac{1}{2 \pi i} \int \frac{\G(-s-1/2) \, (\G(s+1))^2}{\G(1-s)} 
\, x^{-2s} \, ds \, , \]
in which the contour of integration on the $s$-plane is between the poles
of $\G(-s-1/2)$ and $\G(s+1)$. By using the properties of the $\G$-functions,
one can also write
\[ u(x) = - \frac{1}{2 \pi i} \int 
\frac{s^2 \, \G(s) \, \tan{\pi s}}{\G(s+3/2)} \, x^{-2s} ds \, , \]
and note that $\G(-s-1/2)$ has poles at $s=-1/2, 1/2, 3/2, \dots$  
and $\G(s+1)$ has poles at $s= -1, -2, -3, \dots$.

\begin{acknowledgments}

This work is partially supported by the Scientific and Technological Research
Council of Turkey (T{\"U}B\.{I}TAK). B.T. is also partially supported by
the Turkish Academy of Sciences (T{\"U}BA) and by the T{\"U}B\.{I}TAK 
Kariyer Grant No. 104T177. 

\end{acknowledgments}

\end{document}